\documentclass[eat,twocolumn]{jmlr}

\usepackage{longtable}

\usepackage{booktabs}
\usepackage[load-configurations=version-1]{siunitx} 

\usepackage{amsmath,amssymb}

\usepackage{enumitem}
\usepackage{xcolor}

\usepackage{cite}
\usepackage{natbib}

\usepackage[normalem]{ulem}
\useunder{\uline}{\ul}{}
\usepackage{hyperref}

\theorembodyfont{\upshape}
\theoremheaderfont{\scshape}
\theorempostheader{:}
\theoremsep{\newline}

\jmlrvolume{}
\firstpageno{1}
\editors{}

\jmlryear{2022}
\jmlrworkshop{Machine Learning for Health (ML4H) 2022}

\title[Improving Computed Tomography (CT) Reconstruction via 3D Shape Induction]{Improving Computed Tomography (CT) Reconstruction via 3D Shape Induction}

  \author{\Name{Elena Sizikova} \Email{es5223@nyu.edu}\\
  \addr New York University
  \AND
  \Name{Xu Cao} \Email{xc2057@nyu.edu}\\
  \addr New York University
  \AND
  \Name{Ashia Lewis} \Email{atlewis5@crimson.ua.edu}\\
  \addr The University of Alabama
  \AND
  \Name{Kenny Moise} \Email{kenny.moise@uniq.edu}\\
  \addr Université Quisqueya 
  \AND
  \Name{Megan Coffee} \Email{megan.coffee@nyulangone.org}\\
  \addr NYU Grossman School of Medicine 
 }

\begin{document}

\maketitle

\begin{abstract}
Chest computed tomography (CT) imaging adds valuable insight in the diagnosis and management of pulmonary infectious diseases, like tuberculosis (TB). However, due to the cost and resource limitations, only X-ray images may be available for initial diagnosis or follow up comparison imaging during treatment. Due to their projective nature, X-rays images may be more difficult to interpret by clinicians. The lack of publicly available paired X-ray and CT image datasets makes it challenging to train a 3D reconstruction model. In addition, Chest X-ray radiology may rely on different device modalities with varying image quality and there may be variation in underlying population disease spectrum that creates diversity in inputs. We propose shape induction, that is, learning the shape of 3D CT from X-ray without CT supervision, as a novel technique to incorporate realistic X-ray distributions during training of a reconstruction model. Our experiments demonstrate that this process improves both the perceptual quality of generated CT and the accuracy of down-stream classification of pulmonary infectious diseases.
\end{abstract}
\begin{keywords}
CT; 3D Reconstruction; Analysis of Pulmonary Diseases
\end{keywords}

\section{Introduction}
This paper tackles the problem of 3D reconstruction of CT from X-ray images. The ability to visualize 3D internal anatomy allows medical professionals to better assess abnormal pulmonary findings, including those associated with infectious diseases, such as COVID-19 or TB. In particular, TB is a global infectious disease with 1.7 billion infected worldwide and over 1.5 million deaths a year, and is currently (as of 2022) second only to COVID-19 as an infectious cause of death annually.

In many rural and under-served populations, access to radiology is limited. X-rays, which are more accessible, do not provide the nuanced insight multiple slices of imaging a CT scanner provides. A single X-ray is cheaper and easier to obtain and hence may be the only accessible option, besides ultrasound. Even when CTs are  available, follow up with repeat CTs is not practical and would expose patients to unnecessary radiation. However, X-rays contain only a projected, noisy subset of the visual information present in CT, making it more challenging to identify certain pathological features. Additionally, there may not be enough radiologists and other clinicians present to interpret either the X-ray or the CT images in resource-limited settings. The lack of paired chest X-ray and CT datasets, the diversity of patient findings and image quality limitations make the task of learning generative CT models from X-ray inputs a particular challenge. On the other hand, for under-served populations, both an automatic technique for chest CT prediction and for diagnosis of pulmonary infectious disease would improve the quality and timely delivery of healthcare~\citep{lewis2021improving}. 

\begin{figure*}[t]
    \centering
    \includegraphics[width=0.6\linewidth]{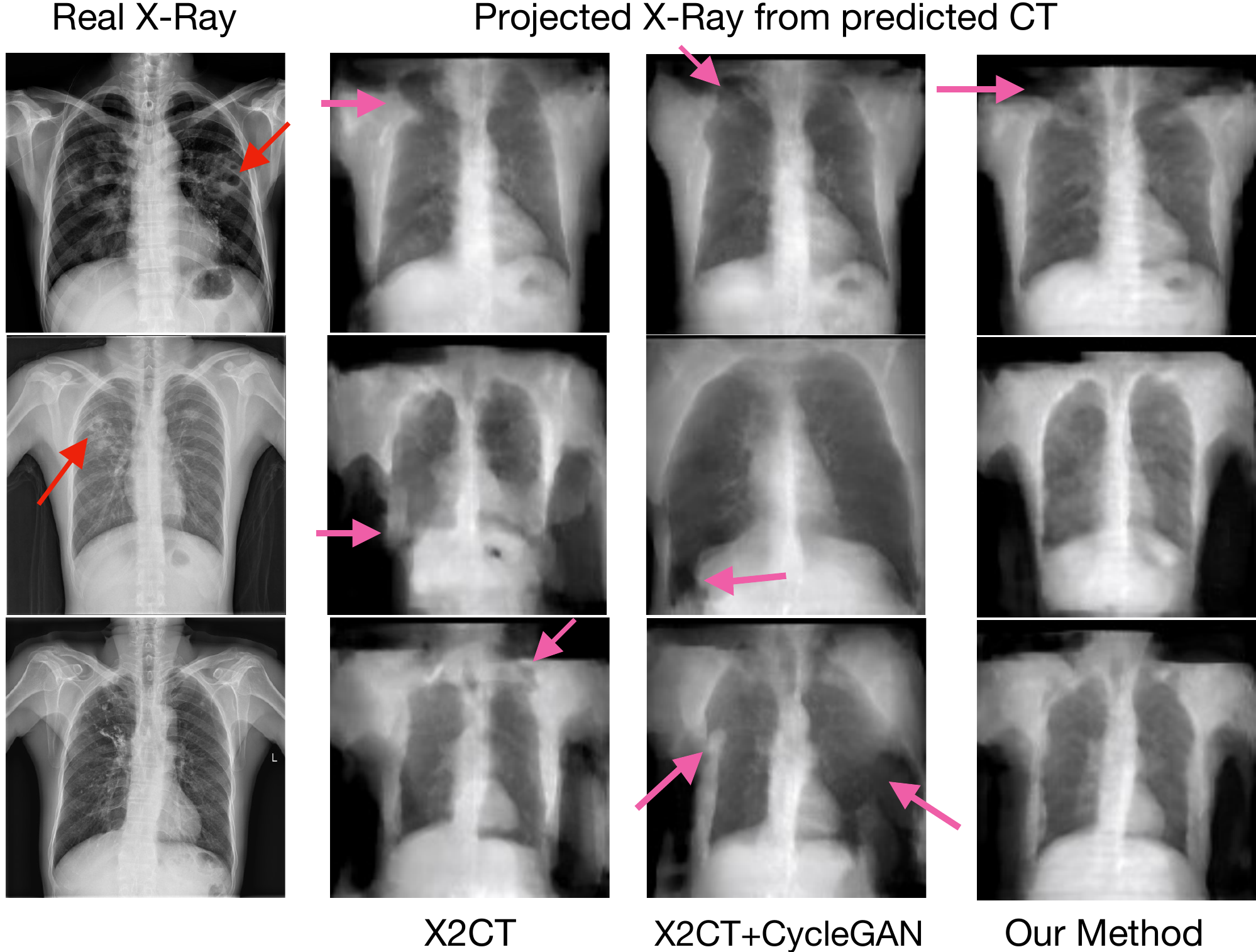}
    \caption{Visualization of real X-rays from the TBX11K~\citep{liu2020rethinking} dataset, as well as the corresponding original and fine-tuned projections from predicted CT images. We find that our proposed method better captures anatomical details, such as the shape of the lung and the pleural and thoracic integrity, avoiding gaps in the thoracic wall and pleural membrane surrounding the lungs seen on the projected X-rays and their corresponding generated CTs (not shown). The red arrow identifies the pulmonary abnormalities in the lung fields bilaterally which likely corresponds to TB and is lost in the projected images. The pink arrows show where there are gaps in the thoracic wall and pleura, which normally separate the lungs from the outside world, but here the lungs are imaged as continuous with the space outside the body. These unrealistic findings, not seen in the real X-rays, were less common in our method.}
    \label{fig:sample_projections}
\end{figure*}

In this work, we build upon the system proposed in X2CT~\citep{ying2019x2ct} in several ways. In particular, this system trains on pairs of real CT images and X-rays generated synthetically using the digitally reconstructed radiograph (DRR) technology~\citep{milickovic2000ct}, due to the lack of available paired training data. To bridge the gap between real and synthetic input data distributions, CycleGAN~\citep{zhu2017unpaired}, an unpaired image translation pre-processing, is used. We observe that this image pre-processing step introduces noise and degrades the quality of input X-rays. In addition, available real X-ray images are not used to improve the expressiveness of the CT generative model. We therefore propose to use shape induction~\citep{gadelha20173d} as a strategy to address these limitations. Inspired by average intensity projection (AIP) and ray summation~\citep{dalrymple2005introduction}, we introduce a shape induction step: when the generative model is presented with an example of an X-ray image without a corresponding CT, the system evaluates whether the projected X-ray obtained from a predicted CT image matches the input X-ray (see Figure~\ref{fig:sample_projections}). This allows us to incorporate any input X-ray datasets directly during training of the CT generative model. We find that the CT images created by our system are more predictive of disease and better capture diversity in clinical radiology findings. Through this work, we highlight the challenge of capturing image variability in automatic techniques for generative modeling and classification in medical imaging, and show how unsupervised shape induction can benefit healthcare accessibility in low-resource healthcare settings. 

CT and X-rays use the same imaging technique but different actual machines. CTs are standardly only available in high resource tertiary medical settings; X-rays can be available in rural, low resource settings, cost less, create less radiation exposure, and can be done at bedside. CT is preferred by clinicians because it provides more detailed imaging which is intuitively comprehensible and mirrors the lungs anatomically. Moving between successive image slices when reading a CT, one can show the cavities and nodules that would be obscured in a single image where findings would overlap. By reconstructing CTs from X-rays, clinicians can provide more targeted clinical care, train with this imaging technology otherwise potentially inaccessible in low resource settings, and improve their understanding of TB, benefiting their patients overall. On the other hand, automatic classification streamlines screening of many images, saving clinician/patient wait times and reduces risk of missing infectious TB cases. Our contributions are as follows. (1) We propose to train an X-ray-to-CT reconstruction model with shape induction, which allows learning from X-ray images only, without CT annotation. (2) We show that the CT images from a model trained with shape induction more accurately classify TB by up to 6.46\% and better capture diversity in the underlying patient pulmonary findings, as judged by a medical professional\footnote{Our code is available at:\url{https://github.com/esizikova/medsynth_public}}.

\section{Approach}
Let $B^{CT} = \{b^{CT}_1,b^{CT}_2,\ldots,b^{CT}_n\}$ be a dataset of real CT images. Using the digitally reconstructed radiograph (DRR) mapping $g$, one can obtain a set of realistic, synthetic images $\hat{G} = \{ \hat{g}_1, \hat{g}_2,\ldots ,\hat{g}_n \}$ such that $\hat{g}_i = g(b^{CT}_i)$ for $i=1,2,\ldots n$. Let $G$ and $D$ be a generator and a discriminator, respectively, that map the distribution of X-ray images to their corresponding CT image counterparts. In X2CT~\citep{ying2019x2ct}, $G$ and $D$ are trained using the LSGAN~\citep{mao2017least} training process. Specifically, $G$ is optimized using a combination of the following objectives. 

The first objective is the adversarial loss $L_{LSGAN} := \frac{1}{2} E [(D(G(x)|x)-1)^2 ]$, where $x\sim d_{\hat{G}}$ and $d_{\hat{G}}$ is the distribution of X-rays generated using DRR. The $L_{LSGAN}$ objective encourages the output CT to be realistic. The second objective is the reconstruction loss $L_{re} := \| y-G(x)\|^2_2$, which measures the MSE between each GT CT $b^{CT}_i$ and its estimate $\hat{b}^{CT}_i = G(\hat{g}_i)$. The third objective is the projection loss $L_{pl} := \frac{1}{3}\sum_{i=1,2,3} \|(Proj_{i}(y) - Proj_{i}(G(x))\|_1$, where $Proj_{1,2,3}$ are the axial, coronal, and sagittal plane projections, respectively, and $\|\cdot\|_1$ is the $L_1$ distance. $L_{pl}$ encourages the 2D projections of $b^{CT}_i$, the GT CT,  and $G(\hat{g}_i)$, the predicted CT, to match.

In practice, the real X-ray distribution $p_{data}$ may not match the DRR X-ray distribution $d_{\hat{G}}$. Therefore, \citep{ying2019x2ct} proposes to use an image translation approach to map each realistic input $x\sim p_{data}$ to its DRR-style version $\hat{x} = s(x)$, where $s$ is a learnt function. Therefore, the CT corresponding to $x$ will be approximated using $G(\hat{x})$, making the assumption that $x\approx \hat{x}$. In our work, we argue that $s$ may lose information about $x$, i.e., $x\not\approx \hat{x}$, and propose to learn $G$ jointly on both $p_{data}$ and $d_{\hat{G}}$, using shape induction, described below. 

\noindent \textbf{Shape Induction} Let $D^{X} = \{d^{X}_1,d^{X}_2,\ldots,d^{X}_m\}\sim p_{data}$ be a dataset of real X-ray images (data points in $B^{CT}$ and $D^{X}$ are independent, and typically collected from different patients). Given an X-ray $x$ and its predicted CT image $G(x)$, the coronal projection $Proj_{1}(G(x))$ is a ray sum estimate, whose appearance is similar to the input X-ray~\citep{dalrymple2005introduction}. Based on this observation, our key idea is to define a new shape induction loss function, $L_{sind} := \|Proj_{2}(G(x)) - x \|^2_2$, where $x\sim p_{data}$. $L_{sind}$ measures, via MSE, whether the appearance of the real X-ray and its projected version from the predicted CT match. Notice that $L_{sind}$ does not require the presence of ground truth $CT$ corresponding to $x$. 

We train $G$ using a combined, weighted total loss function $L_G:= \lambda_1 L_{LSGAN} + \lambda_2 L_{re} + \lambda_3 L_{pl} + \lambda_4 L_{sind}$. $D$ is trained using the standard LSGAN discriminator loss (please see \citep{ying2019x2ct}). 

\section{Experimental Details}
We evaluate our approach on the public dataset TBX11K~\citep{liu2020rethinking}. We use the official data split in \citep{liu2020rethinking} that also includes multiple smaller, public sets~\citep{jaeger2014two,chauhan2014role}. For qualitative performance evaluation, we evaluate on COVID-19-CT-CXR~\citep{peng2020covid} subset that consists of paired X-ray and CT images of patients from COVID-19 articles in the PubMed Central Open Access (PMC-OA).  The images considered in this experiment are pathological examples cropped from COVID-19 articles in the PubMed Central Open Access (PMC-OA). For example, they include annotations, such as figure numbers ``A" and ``B" in the first line of Figure~\ref{fig:qual_comparison}, which our model was not trained on. Due to the nature of the dataset and the image collection process, these images are a difficult test set for our system. Two board-certified medical professionals evaluated and commented on the visual artifacts found in both X-rays and CTs. 

\section{Results}
In this section, we present qualitative and quantitative results of our method. For quantitative evaluation of clinical correctness, we measure if the generated images can predict the presence of TB.

\begin{table*}[htb]
\centering
\resizebox{0.8\textwidth}{!}{
\begin{tabular}{|l|c|c|l|}
\hline
                                                                  & \multicolumn{1}{l|}{X2CT~\citep{ying2019x2ct}} & X2CT+CycleGAN~\citep{zhu2017unpaired} & Our Method            \\ \hline
\begin{tabular}[c]{@{}c@{}}Classification Accuracy\end{tabular} & {\ul 0.925$\pm$0.003}        & 0.882$\pm$0.005  & \textbf{0.939$\pm$0.003} \\ \hline
\begin{tabular}[c]{@{}c@{}}TB Recall\end{tabular}               & {\ul 0.789$\pm$0.008}        & 0.693$\pm$0.038  & \textbf{0.826$\pm$0.006} \\ \hline
\begin{tabular}[c]{@{}c@{}}TB Precision\end{tabular}               &  {\ul 0.915$\pm$0.019}      & 0.866$\pm$0.021  & \textbf{0.934$\pm$0.008} \\ 

\hline
\end{tabular}
}
\caption{Quantitative comparison across different CT generation techniques. Our method (using shape induction) obtains best results. Best result is in bold and the second best is underlined.}
\label{tab:3d_classification}
\end{table*}

\noindent \textbf{Comparison of CT Generation Strategies}
In Table~\ref{tab:3d_classification}, we compare the CT images generated using different training strategies on the task of classification into healthy, sick, and TB. We find that the proposed method trained with shape induction obtains both the highest classification accuracy and the best recall of TB examples. On the other hand, we find that the cycleGAN~\citep{zhu2017unpaired} CT offers less prediction value, with lower classification scores and lower TB recall rates. This is because cycleGAN~\citep{zhu2017unpaired} limits the expressiveness of X-rays, normalizing clinical abnormalities, and adds noise (see Appendix).

\begin{figure*}[htb]
    \centering
    \includegraphics[width=0.8\textwidth]{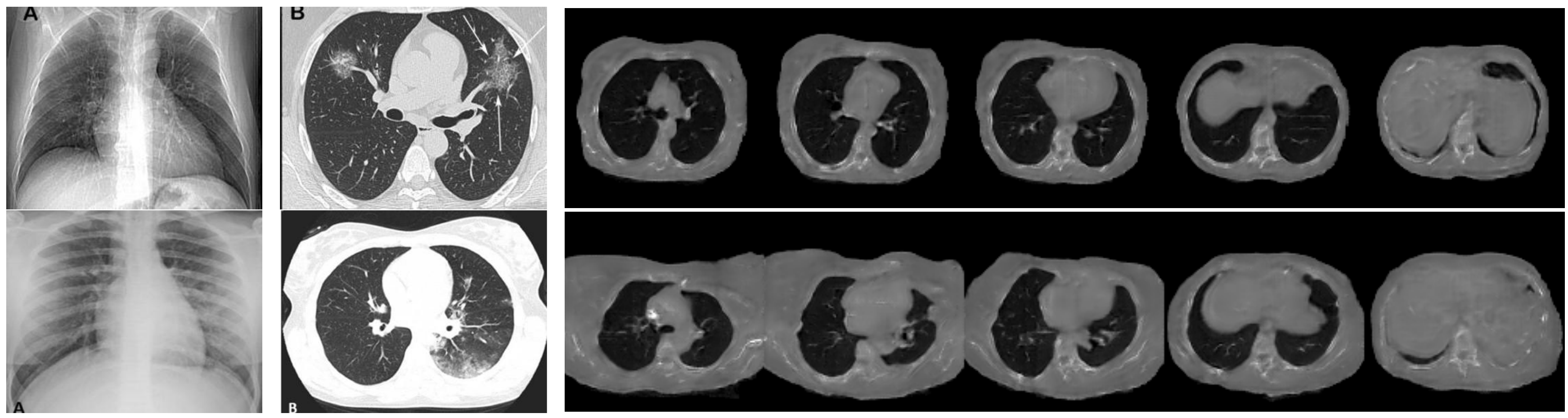}
    \caption{Qualitative comparison of CT generated from ground truth X-rays and corresponding real CT images obtained from the COVID-19-CT-CXR~\citep{peng2020covid} dataset (images from \citep{schmitt2020covid,musolino2020lung}). While all three methods exhibit significant artifacts, we find that the the proposed technique is better at a) capturing some of the abnormalities associated with disease without over-sanitizing and b) avoiding anatomical abnormalities (thoracic wall defects) not consistent with Chest X-rays, as seen with other methods.} 
    \label{fig:qual_comparison}
    
\end{figure*}

\noindent \textbf{Qualitative Comparison of Real and Predicted CT Images}
In Figure~\ref{fig:qual_comparison}, we visualize the input X-rays, ground truth and predicted CT images from the COVID-19-CT-CXR~\citep{peng2020covid} dataset. The images in this dataset are obtained from scraping public articles and depict pulmonary findings of severe COVID-19 and other respiratory diseases, therefore, presenting a particularly challenging case for our algorithm. Compared to the X2CT~\citep{ying2019x2ct} and X2CT+CycleGAN~\citep{zhu2017unpaired}, our model captures more anatomic details of the CT images.

\section{Limitations}
The model described in this paper presents some limitations. In many instances, it failed to capture basic anatomic features, such as the integrity of the chest wall. It developed aberrant anatomic structures such as three lungs on CT in other cases. In other instances, while capturing basic anatomy, it failed to detect pulmonary findings on CT while they are present on X-rays. Therefore, the model did not create the ideal CT for clinical use with every output. When comparing classification based on X-rays and generated CT, we did not observe a consistent improvement in classification performance using the generated CT. We assess our approach qualitatively and via a proxy task (classification), and while evaluation using image reconstruction metrics such as SSIM/L2 would be more preferable, this approach would require paired X-ray and CT data, which is unavailable for our application. Further evaluation is necessary to ensure that our evaluation is not susceptible to shortcut learning, for example, that the decision of classification model decisions are based on reconstructions of clinically relevant pathologies.

The model has also been run on a relatively low number of input X-rays which are not able to encompass the full diversity of findings in pulmonary TB, let alone the wider range of disease in pulmonary images. Moreover, the datasets used here did not include the lower-quality images often relied upon in resource-limited settings, especially where digital X-rays are not available and plain films are used. These databases are also not representative of the populations where TB is more common. In areas where TB is more frequent, patients present more commonly with a primary infection, more frequently at a younger age, accessing care with more advanced disease, than in populations where the disease is less common and often due to reactivation of past disease in older adults, and may have different lung findings, and can access care and imaging earlier before more substantial lung pathology develops. HIV can lead to TB manifesting with few radiographic findings and images would need to be compared to a more varied backdrop of other pulmonary disease findings as a wide range of opportunistic infections are common in those with HIV, but not those who are HIV negative.

\section{Conclusion and Future Work}
In this work, we present an approach for training an automatic chest CT reconstruction algorithm with X-ray only. We augment existing model training on DRR-generated X-ray and CT pairs with a shape induction loss, allowing the model to learn from only real input X-rays. This approach allows learning the variability of real X-ray images and directly  incorporating it into the training of the CT generation model. The ability to learn rich distributions from real X-rays is particularly important for practical applications where there is a need to adapt to different imaging sensor types and anatomies. 

In future work, it will be important to devise techniques to constrain the model to generate more realistic outputs and ensure the model is not generating sanitized CT images without the abnormalities representing disease processes. Clearly representing abnormal lung findings would help clinicians identify the type of illness (such as TB) and features of this illness which may affect management (may need fluid drainage, may have lifelong damage). At the same time, it is important that the CT images do not misrepresent findings and show gross and erroneous abnormalities (e.g., chest wall gaps, additional lungs) not consistent with these patients being alive and hence would decrease the trust in the tools by clinicians using the tool. In addition, it would be extremely valuable to design a system that would focus on capturing and identifying distinct lung findings indicative of TB, including cavities, apical capping, micronodules of miliary TB, as well as features often shared with other diseases, like infiltrates and fluid collecting in pleural space (pleural effusion). Further model development will require expansion of existing databases to more robustly represent the TB populations where disease is often more severe at time of imaging, and may include different types of findings seen in younger patients with initial exposures to TB. Overall, this work is a step towards developing a tool which can classify types of pulmonary disease, including TB and its specific subtypes, as well as provide clinicians with CT imaging after obtaining only chest X-rays.

\bibliography{pmlr-sample}
\clearpage
\section*{Appendix}

\subsection{Related Work}

3D reconstruction is an important ongoing area of research in medical image analysis~\citep{ying2019x2ct}. \citep{caponetti19903d} is one of the earliest attempts at capturing bone structure based on back-lighting projections. 
Close to our work, \citep{henzler2018single} and \citep{ying2019x2ct} proposed to reconstruct CT from X-rays using neural networks. Due to the lack of availability of paired 2D-3D supervision, both of these works train on synthetic X-ray inputs and cannot directly update the generative CT model based on real X-rays. \citep{ying2019x2ct,kasten2020end} rely on unsupervised style transfer (CycleGAN~\citep{zhu2017unpaired}) to transform between real and synthetic X-ray inputs. On the other hand, unsupervised 3D generative modeling has been addressed using various approaches in image-based reconstruction
~\citep{henzler2019escaping,gadelha20173d}. \citep{chen2021equivariant} used image equivariance as a self-supervision task for image reconstruction. In particular, \citet{gadelha20173d} proposed a GAN-based shape induction approach trained from a set of shape projections only. 

The goal of computer-aided diagnosis approaches is to speed up healthcare delivery, while reducing time and costs associated with analysis of radiologic findings~\citep{doi2007computer}. 
CT often offers additional information and clearer views of pulmonary findings
~\citep{burrill2007tuberculosis,alkabab2018performance}. Similar to X-rays, automatic CT images have been used for analysis of COVID-19~\citep{pham2020comprehensive}, lung cancer~\citep{kuruvilla2014lung} and TB~\citep{zunair2020uniformizing}. Furthermore, \citep{maghdid2021diagnosing} combined X-rays and CT to improve disease classification accuracy. Fusing information across different sources of information
~\citep{zhu2021mvc} has also shown superior performance over that of individual modalities.

\subsection{Additional Experimental Details}
\noindent \textbf{Datasets} Since no annotations are available for the test images in TBX11K~\citep{liu2020rethinking}, we split the original validation $1:1$ into validation set and testing set. Therefore, our testing set contains 6,888 training images, 1,044 validation images, and 1,044 test images. For the X2CT~\citep{ying2019x2ct}, we use the official implementation trained on the LIDC-IDRI dataset~\citep{armato2011lung} with 1,018 chest CT scans. 

\begin{figure}[tb]
    \centering
    \includegraphics[width=1.0\linewidth]{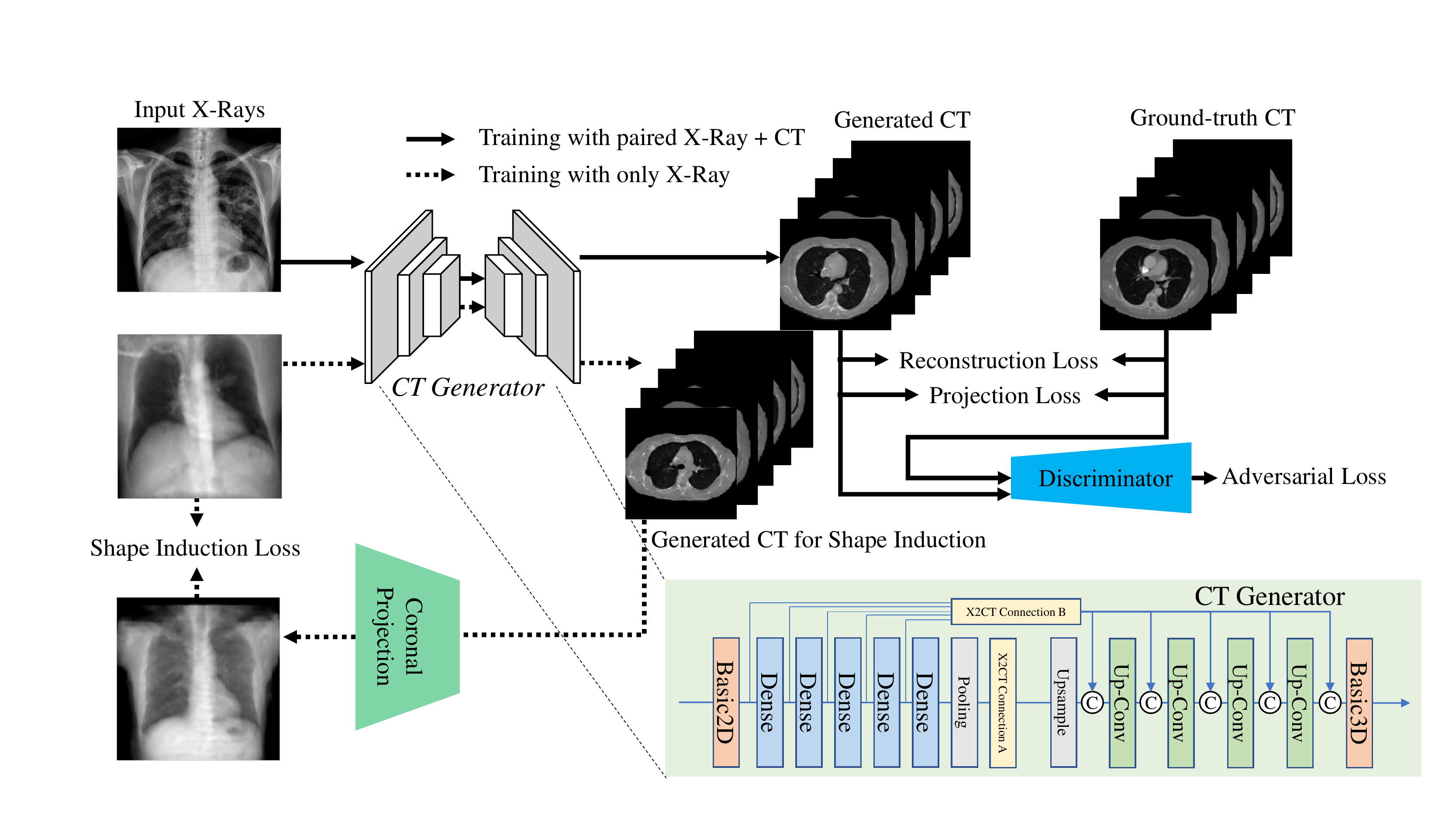}
    \caption{Overview of our system. We introduce a shape induction loss to facilitate finetuning a CT reconstruction system on datasets with only X-ray images.}
    \label{fig:overview}
\end{figure}

\paragraph{Architectures} An overview of the components of our system can be seen in Figure~\ref{fig:overview}. For CT generation, we use the single-view X2CT~\citep{ying2019x2ct} architecture. In all experiments, $\lambda_{1,2,3}=0.1$, which is suggested by~\citep{ying2019x2ct}. We conduct ablation studies for selecting $\lambda_4$. 

\begin{table*}[t]
\centering
\resizebox{1.0\textwidth}{!}{
\begin{tabular}{|l|c|c|c|c|c|}
\hline
  &
  $\lambda_4=0.0$ & $\lambda_4=0.1$ & $\lambda_4=1.0$ & $\lambda_4=10.0$ & $\lambda_4=100.0$ \\ \hline
Classification Accuracy & 0.925$\pm$0.003 & 0.930$\pm$0.004 & 0.937$\pm$0.003 & \textbf{0.939$\pm$0.003} & 0.936$\pm$0.013 \\ \hline
TB Recall & 0.789$\pm$0.008  & 0.806$\pm$0.016 & 0.817$\pm$0.034 & \textbf{0.826$\pm$0.006} & 0.821$\pm$0.052\\ \hline
TB Precision & 0.915$\pm$0.019 & 0.916$\pm$0.014 & 0.925$\pm$0.016 & 0.934$\pm$0.008 & \textbf{0.940$\pm$0.024}\\ \hline
\end{tabular}
}
\caption{Ablation study of $\lambda_4$, weight influence of shape induction loss.}
\label{tab:abl_hyper}
\end{table*}

\begin{figure*}[]
    \centering
    \includegraphics[width=0.7\textwidth]{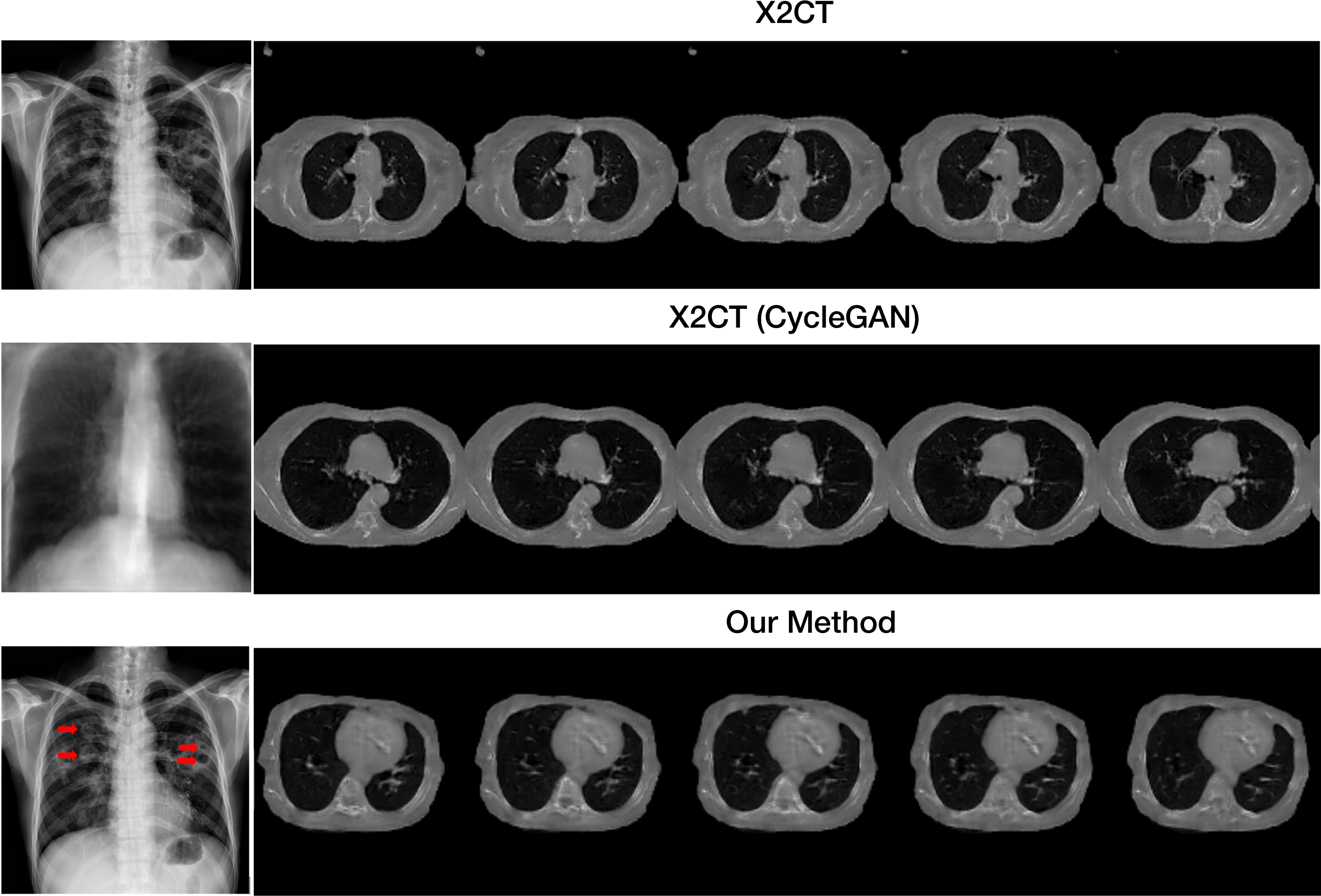}

    \caption{Comparison of CTs generated using different strategies. }
    \label{fig:3d_comparison_qual}
    
\end{figure*}

For X-ray classification, we use the TB classification model~\citep{DUONG2021115519} using the ViT-S with $16 \times 16$ patch size~\citep{dosovitskiy2020image}. Similar to \citep{DUONG2021115519}, we use a classification head that maps transformer features to $K=3$ classes. For CT classification, we use the 3D-CNN model previously proposed in \citep{zunair2020uniformizing} for TB classification from CT. The performance of each model is averaged over three training runs. All experiments are performed in PyTorch using a single NVIDIA Quadro RTX8000 GPU with 100GB CPU memory cap.

\noindent \textbf{Training Details} For shape induction, we pre-train $D$ and $G$ for $n^1_e = 90$ epochs, then alternate fine-tuning on $D^X$ and $B^{CT}$ for $n^2_e = 10$ epochs. The baseline model (single-view X2CT~\citep{ying2019x2ct}) is trained on $B^{CT}$ only for $n_e = 100$ epochs. This model consists of a 2D encoder and a 3D decoder CNN with skip connections between the encoder and decoder. A 3D PatchGAN~\citep{isola2017image} discriminator is used for adversarial training. The X-Ray classification model is trained for $n=50$ epochs, and the best model according to the validation set is used for testing.  The model consists of four convolutional blocks, followed by a global average pooling (GAP) layer and two fully connected layers, with number of outputs 512 and number of classes $K=3$, respectively, with a dropout layer in between. The model is also trained for $n=50$ epochs, and the best model, based validation, is used for testing. The CT reconstruction takes about 22 hours to train, the classification model takes about 5 hours (CT) and 1 hour (X-Ray). 

\noindent \textbf{Analysis of Shape Induction Influence}
Based on the experiments in Table~\ref{tab:abl_hyper}, we find that using larger shape induction weight $\lambda_4$ improves the generation of 3D CT. We compare the classification model using 3D CT generated by $\lambda_4=0$ (original X2CT mode), and $\lambda_4=0.1, 1.0, 10.0, 100.0$, and find that $\lambda_4=10.0$ achieves most accurate downstream CT classification model. We perform the ablation study and selected the optimal value of $\lambda_4$ based on the validation set ($lambda_4=10.0$). Results in Table~\ref{tab:abl_hyper} reflect performance on the test set.

\paragraph{Visual Comparison of Generated CT}
We report a visual comparison of CT generated using different strategies in Figure~\ref{fig:3d_comparison_qual}. CTs that use CycleGAN~\citep{zhu2017unpaired} X-rays often exhibit less artifacts as the input X-rays are closer to the training distribution. On the other hand, the resulting CT images are less expressive and do not capture the abnormalities seen in the Chest X-rays which would be associated with TB. By using shape induction, we are able to maintain the generated image variability while matching the distributions in input X-rays. Here, red arrows pointing to pulmonary abnormalities are seen on the Chest X-ray and some abnormalities are seen on the CT images in the expected, corresponding location. These expected disease-associated abnormalities were not captured in the CycleGAN approach.

\section{Acknowledgments}
\noindent ES was supported by the Moore-Sloan Data Science Environment initiative (funded by the Alfred P. Sloan Foundation and the Gordon and Betty Moore Foundation) through the NYU Center for Data Science. AL was supported by the NYU Center for Data Science (CDS) Undergraduate Research Program (CURP) in partnership with the National Society of Black Physicists (NSBP). ES, XC, MC, KM were supported from by the NYU Research Catalyst Prize. We thank anonymous reviewers for their feedback.





\end{document}